\documentstyle[12pt]{article}
\normalsize

\input mydef.tex

\let\oldtheequation=\theequation
\def\doteqs#1{\setcounter{equation}{0}
            \def\theequation{{#1}.\oldtheequation}}

\newcounter{sxn}
\def\sx#1{\addtocounter{sxn}{1} \bigskip\medskip \goodbreak \noindent{\large\bf
\centerline{\thesxn.~~#1}} \nobreak \medskip}
\def\sxn#1{\sx{#1} \doteqs{\thesxn}}

\newcounter{axn}

\def\br{\begin{thebibliography}{abc}}
\def\er{\end{thebibliography}}

\date{}

\begin{document}
\bibliographystyle{unsrt}
\footskip 1.0cm
\thispagestyle{empty}
\setcounter{page}{0}
\begin{flushright}
UAHEP 9114\\
SU-4228-500\\
INFN-NA-IV-91/25\\
DSF-T-91/25\\
G\"oteborg ITP 91-59\\
January 1992
\end{flushright}
\begin{center}{\LARGE POISSON BRACKET ALGEBRA FOR\\
 CHIRAL GROUP ELEMENTS IN\\
  THE WZNW MODEL\\ }
\vspace*{6mm}
{\large   G. Bimonte $^{1,2}$ ,
          P. Salomonson $^{3}$ \\
          A. Simoni $^{2}$ ,
          A. Stern $^{2,4}$ \\ }
\newcommand{\bc}{\begin{center}}
\newcommand{\ec}{\end{center}}
\vspace*{5mm}
 1){\it Department of Physics, Syracuse University,\\
Syracuse, NY 13244-1130, USA}.\\
\vspace*{4mm}
 2){\it Dipartimento di Scienze Fisiche dell' Universit\`a di Napoli,\\
    Mostra d'Oltremare pad. 19, 80125 Napoli, Italy,\\
    and\\
 Istituto Nazionale di Fisica Nucleare, Sezione di Napoli,\\
    Mostra d'Oltremare pad. 19, 80125 Napoli, Italy}.\\
\vspace*{4mm}
 3){\it Institute of Theoretical Physics,\\
S-41296 G\"oteborg, Sweden}.\\
\vspace*{4mm}
 4){\it Department of Physics, University of Alabama, \\
Tuscaloosa, AL 35487.}\ec

\vspace*{5mm}

\normalsize
\centerline{\bf ABSTRACT}
We examine the Wess-Zumino-Novikov-Witten (WZNW) model on a circle and
compute the Poisson bracket algebra for left and right moving
chiral group elements.  Our computations apply for arbitrary
groups and boundary conditions, the latter being
 characterized by the monodromy matrix.  Unlike in previous treatments,
they do not require specifying a
particular parametrization of the group valued fields
in terms of angles spanning the group.
We do however find it necessary to make a gauge
choice, as the chiral group elements are not gauge invariant observables.
(On the other hand,
the quadratic form of the Poisson brackets may be defined independent
of a gauge fixing.)
Gauge invariant observables can be formed from the monodromy matrix and
these observables are seen to commute in the quantum theory.

\newpage
\newcommand{\be}{\begin{equation}}
\newcommand{\ee}{\end{equation}}
\baselineskip=24pt

\sxn{Introduction}

The symplectic structure for the Wess-Zumino-Novikov-Witten
(WZNW) model is well know when written
in terms of its conserved currents.[1-5]  It
corresponds to Kac-Moody algebras and defines a conformal block.
Recently, there has been interest in obtaining the associated algebra for
the left and right moving chiral group elements $u$ and $v$
of the theory.[6-11]  A characteristic feature of
the Poisson brackets found for the chiral group elements
is that they are quadratic in $u$ and $v$. It was then
conjectured that they are the classical analogues of the
quantum exchange
algebras shown for the WZNW model.\cite{rehr,moor}
However, $u$ and $v$ are not gauge invariant observables
in the theory, so the physical meaning of their Poisson brackets and
their associated commutation relations is not immediately transparent.
On the other hand, the quadratic nature of the Poisson brackets
can be defined independent of any choice of gauge.  We explain this
below.

The fundamental fields $g$ of the model take values in an $n$
dimensional Lie group $G$.  We shall restrict our attention to $G$ being
a compact semisimple group.  The dynamics is given by the WZNW action
\begin{equation}
 S =  -{\kappa \over{16\pi}}
 \int_{\partial {\cal D}} dxdt  \;Tr(g^{-1}\partial_\mu g
g^{-1}\partial^\mu g) \quad +\quad
{\kappa \over{24\pi}} \int_{\cal D} Tr(g^{-1}dg)^3\quad,
\end{equation}
where ${\cal D}$ is a disc $\times$ time, and we have already made a
specific choice for the coefficients of the two terms in eq. (1.1).
The equations of motion can be written on
the boundary $\partial {\cal D}$ of ${\cal D}$ and hence dynamics
for the system takes place at $\partial {\cal D}$.  We parametrize the
boundary by $x$ and $t$, $t$ being
the time variable and $0\leq x\leq \ell$.  The equations of motion are
\begin{equation}
\partial_+ (g^{-1}\partial_- g )=0\quad,
\end{equation}
where $x_{\pm}=t\pm x$.
The general solutions of the equations of motion
are of the form $g(x,t)=u(x_+)v(x_-)$.  $u$ and $v$ denote the left and
right chiral group elements, respectively.  They take values in $G$.

We assume that $g$ is singlevalued on the circle, $g(x+\ell,t)=g(x,t)$.
Although $g(x,t)$ is well defined on the circle,
 $u$ and $v$ need not be.  Rather,
\begin{equation}
u(x+\ell)=u(x){\cal M} \quad{\rm and}\qquad
  v(x+\ell)={\cal M}v(x)\quad    .
\end{equation}
${\cal M}$ denotes the monodromy and it takes values in $G$.
Furthermore there are extra (gauge) degrees of freedom in $u$
and $v$, not present in the field degrees of freedom $g(x,t)$.
This is since the transformation
\begin{equation}
u(x)\rightarrow u(x)h \quad{\rm and}\quad
  v(x)\rightarrow h^{-1}v(x)
\end{equation}
leaves $g(x,t)$ invariant.  Here $h\in G$ is independent of $x$.
 The monodromy also is not gauge invariant, as it undergoes
a similarity transformation under (1.4),
\begin{equation}
{\cal M}\rightarrow h^{-1} {\cal M} h.
\end{equation}
${\cal M}$ is thus mapped to an adjoint orbit under the action of
the gauge group, and adjoint invariants can be formed from ${\cal M}$
which have a gauge invariant meaning.  We will find that the Poisson
brackets of these invariants with themselves are zero, and thus
the corresponding quantum operators can be simultaneously diagonalized.

The Poisson brackets between chiral group elements have been
found in the literature to have the following form:[6-11]
\begin{equation}
\{u(x){}^\otimes_, u(y)\} = u(x)T_\rho\otimes u(y)T_\sigma\;
{\cal F}_{\rho\sigma}(x,y)\quad,
\end{equation}
\begin{equation}
\{u(x){}^\otimes_, v(y)\} = u(x)T_\rho\otimes T_\sigma v(y)\;
{\cal G}_{\rho\sigma}(x,y)\quad,
\end{equation}
\begin{equation}
\{v(x){}^\otimes_, v(y)\} = T_\rho v(x)\otimes T_\sigma v(y)\;
{\cal H}_{\rho\sigma}(x,y)\quad,
\end{equation}
where $\{T_\alpha$, $\alpha=1,2,...,n\}$ correspond to a complete set
of Lie group generators.  In terms of matrix indices $i, j, k,...$,
 the brackets
in eqs. (1.6-8) are defined by
$\{u(x){}^\otimes_, u(y)\}_{ik,jm} = $
$\{u(x)_{ij}, u(y)_{km}\}.$
${\cal F}$ ,${\cal G}$ and ${\cal H}$ denote $n \times n$
matrices which are functions of two variables $x$ and $y$.
They have the following properties:

${ \it i)}$ Antisymmetry of the Poisson brackets requires that
${\cal F}$ and ${\cal H}$ satisfy
\begin{equation}
{\cal F}^T(x,y)=-{\cal F}(y,x)\quad {\rm and } \quad
{\cal H}^T(x,y)=-{\cal H}(y,x)\quad.
\end{equation}

${\it ii)}$ Since $u$ and $v$ are not gauge invariant, neither are
the matrices ${\cal F}$, ${\cal G}$ and ${\cal H}$.  Let $\Phi_\alpha$
be the generators of gauge transformations (1.4) in phase space.
They should satisfy the Poisson brackets
\begin{equation}
\{\Phi_\alpha, u(x)\} =-i u(x)T_\alpha\quad,
\end{equation}
\begin{equation}
\{\Phi_\alpha, v(x)\} = iT_\alpha v(x)\quad.
\end{equation}
In the Dirac-Bergmann theory of constrained Hamiltonian systems
we expect that
$\Phi_\alpha\approx K_\alpha$, $K_\alpha$ being c-numbers,
are first class constraints.
Now consider the Jacobi identity involving repeated Poisson brackets of
$\Phi_\alpha$, $u(x)$ and $u(y)$.  Upon applying (1.6), we find
\begin{equation}
\{\Phi_\alpha, {\cal F}_{\rho\sigma}(x,y)\} =  c_{\alpha\rho\lambda}
{\cal F}_{\lambda\sigma}(x,y) +c_{\alpha\sigma\lambda}
{\cal F}_{\rho\lambda}(x,y)   \quad,
\end{equation}
where $c_{\mu\nu\lambda}$ are totally antisymmetric structure constants,
ie. $ [T_\mu,T_\nu ]= i c_{\mu\nu\lambda} T_\lambda$.
Analogous relations hold for ${\cal G}$ and ${\cal H}$,
\begin{equation}
\{\Phi_\alpha, {\cal G}_{\rho\sigma}(x,y)\} = c_{\alpha\rho\lambda}
{\cal G}_{\lambda\sigma}(x,y) +c_{\alpha\sigma\lambda}
{\cal G}_{\rho\lambda}(x,y)   \quad,
\end{equation}
\begin{equation}
\{\Phi_\alpha, {\cal H}_{\rho\sigma}(x,y)\} = c_{\alpha\rho\lambda}
{\cal H}_{\lambda\sigma}(x,y) +c_{\alpha\sigma\lambda}
{\cal H}_{\rho\lambda}(x,y)   \quad.
\end{equation}
Eqs. (1.12-14) show how
${\cal F}$, ${\cal G}$ and ${\cal H}$ transform under infinitesimal
gauge transformations.  Under finite gauge transformations (1.4),
\begin{equation}
{\cal F}\rightarrow ad \; h\;{\cal F}\;(ad\; h)^{-1},\quad
{\cal G}\rightarrow ad \; h\;{\cal G}\;(ad\; h)^{-1}\quad
{\rm and}\quad {\cal H}\rightarrow ad \; h\;{\cal H}\;(ad\; h)^{-1},
\end{equation}
where $ad\;h$ denotes the adjoint representation of $h$.
In other words, the form of the Poisson brackets given in eqs.
(1.6-8) is unchanged
under (1.4), provided ${\cal F}$, ${\cal G}$ and ${\cal H}$ undergo the
transformations (1.15).

${\it iii)}\;{\cal F}$, ${\cal G}$ and ${\cal H}$ have nonzero Poisson
brackets with the chiral group elements.  For example, to compute the
bracket between  ${\cal F}$ and $u(x)$, we can study the Jacobi identity
$$\{\{u(x)_{ij},u(y)_{km}\},u(z)_{np}\}
\;+\;\{\{u(y)_{km}, u(z)_{np}\}, u(x)_{ij}\} $$
\begin{equation}
\qquad +\;\{\{u(z)_{np}, u(x)_{ij}\}, u(y)_{km}\}\;=\;0 \quad,
\end{equation}
which we shall assume is valid.
Applying the Poisson bracket (1.6) twice we get
$$\{\{u(x){}^\otimes_,u(y)\}{}^\otimes_,u(z)\}=
u(x)T_\rho\otimes u(y)T_\sigma\;
\otimes\{{\cal F}_{\rho\sigma}(x,y),u(z)\}\qquad \qquad$$
$$+\;\Bigl(u(x)T_\mu T_\rho\otimes u(y)T_\sigma\;
{\cal F}_{\mu\nu}(x,z)
\;+\;u(x)T_\rho\otimes u(y)T_\mu T_\sigma\;
{\cal F}_{\mu\nu}(y,z)\Bigr)\;
\otimes\; u(z)T_\nu \;
{\cal F}_{\rho\sigma}(x,y)\quad, $$
corresponding to the first term in eq. (1.16).   The Jacobi identity
can then be rewritten
$$u(x)T_\rho\otimes u(y)T_\sigma\;
\otimes\Big(\{{\cal F}_{\rho\sigma}(x,y),u(z)\}
\;+\;ic_{\mu\nu\lambda}u(z)T_\lambda\;
{\cal F}_{\nu\rho}(z,x)\;
{\cal F}_{\mu\sigma}(z,y)\Bigr)\quad.$$
\begin{equation}
\qquad + \;cyclic\;\; terms\quad =\quad 0\quad.
\end{equation}
The general solution to this equation is
\begin{equation}
\{{\cal F}_{\rho\sigma}(x,y),u(z)\}
\;=\;-\;ic_{\mu\nu\lambda}u(z)T_\lambda  \;
\Big({\cal F}_{\nu\rho}(z,x)  \;
{\cal F}_{\mu\sigma}(z,y)
\;+\;\alpha\; \delta_{\mu\rho}\delta_{\nu\sigma}\Bigr)\quad.
\end{equation}
Here $\alpha$ is a constant on the circle.  Furthermore, it
is gauge invariant, ie.
\begin{equation}
\{ \Phi_\beta,\alpha \} = 0  \quad.
\end{equation}
To derive eq. (1.19) take the Poisson bracket of
eq. (1.18) with $\Phi_\beta$ and once again apply the Jacobi identity.

In this article we compute the functions
${\cal F}_{\rho\sigma}(x,y)$, ${\cal G}_{\rho\sigma}(x,y)$ and
${\cal H}_{\rho\sigma}(x,y)$ explicitly (up to a constant matrix).
This requires making a gauge choice, as the functions are not gauge
invariant.  As in ref. \cite{godd}, our method is to
systematically invert the symplectic two form of the theory.
The latter is given in section 2.  Unlike in
 ref. \cite{godd}, we do not require specifying
a particular group parametrization for $u$ and $v$
in order to do the calculation.

The space $Q$ spanned by the chiral group elements $u(x)$ and $v(x)$
defines a principal $G$ bundle, with
the group valued fields $g(x,t)$ making up the base manifold.
The projection map is obtained from $g(x,t)=u(x_+)v(x_-)$, while
the action of $G$ on the bundle was given in eq. (1.4).
 The symplectic two form of the theory is
nonsingular when evaluated on the tangent space of the base manifold.
On the other hand, its pull back to the bundle $Q$
, which we denote by $\Omega$, is not.  $\Omega$ is degenerate
when acting on the tangent space $TQ$ of $Q$,
the zero or null vectors corresponding
to the vertical tangent directions of the bundle.

In order to invert $\Omega$, we shall restrict it to acting on a certain
subspace of $TQ$, namely the subspace of horizontal vectors.
This procedure is not unique and requires making
the choice of a gauge, although it should be possible to
transform the resulting Poisson brackets to any other gauge using (1.15).
To define our gauge we shall introduce a scalar product
on $TQ$ and demand that horizontal and vertical vectors are
orthogonal with respect to the scalar product.  [Alternatively,
one can avoid introducing a metric on the bundle space
and instead work with the bundle connection.  We, however, will
not pursue such an approach here.]
We illustrate our method
 first for the case of unit monodromy in section 3,
and for the case of arbitrary monodromy in section 4.  We make some
concluding remarks in section 5.

\sxn{The Symplectic Two Form}

In terms of the chiral group elements, the symplectic two form
 $\Omega$ was found \cite{godd} to be
$$\Omega =-{\kappa \over{ 8\pi}}
 \int_0^\ell dx\; Tr\Bigl( u^{-1}\delta u\>
 \partial_x(u^{-1}\delta u)\;+ \;\delta v v^{-1}
 \partial_x(\delta v v^{-1})\Bigr)$$
\begin{equation}
\qquad -\;{\kappa \over{ 8\pi}}
 Tr\Bigl([u^{-1}\delta u](0)\;
 \delta{\cal M} {\cal M}^{-1}
 \; +\;[\delta v v^{-1}](0)\;
 {\cal M}^{-1}\delta{\cal M}
 \Bigr)\quad,
\end{equation}
where $\delta$ denotes an exterior derivative.  It can be
checked that $\Omega$ is \it a) \rm closed and \it b) \rm
invariant under the gauge transformation (1.4) and (1.5).
In general, all terms in eq. (2.1) are necessary to show this.
 To show \it a) \rm we can define
\begin{equation}
\Omega ^L =  -{\kappa \over{ 8\pi}}
 \int_0^\ell dx\; Tr\> u^{-1}\delta u\>
 \partial_x(u^{-1}\delta u)
\end{equation}
and
\begin{equation}
 \Omega ^R = -{\kappa \over{ 8\pi}}
\int_0^\ell dx\; Tr\; \delta v v^{-1}
 \partial_x(\delta v v^{-1})
 \end{equation}
Upon taking their exterior derivative
$$\delta\Omega ^L = -{\kappa \over{ 8\pi}}
 Tr\Bigl([ u^{-1}\delta u]^2 (0)\;
 \delta{\cal M} {\cal M}^{-1}
 \;+\; [ u^{-1}\delta u](0) \;\bigl(
 \delta{\cal M}{\cal M}^{-1}\bigr)^2\;
+\;{1\over 3} \bigl(
 \delta{\cal M}{\cal M}^{-1}\bigr)^3\Bigr)\quad,$$
\begin{equation}
\delta\Omega ^R =
 {\kappa \over{ 8\pi}}
 Tr\Bigl([\delta v v^{-1}]^2 (0)\;
 {\cal M}^{-1} \delta{\cal M}
 \;+\; [\delta v v^{-1}](0)\;\bigl(
 {\cal M}^{-1} \delta{\cal M}\bigr)^2\;
+\;{1\over 3} \bigl(
 \delta{\cal M}{\cal M}^{-1}\bigr)^3\Bigr)\quad,
\end{equation}
we see that the sum of these terms, along with the
 exterior derivative of the second line in eq. (2.1), vanishes,
thus proving \it a)\rm.  The proof of \it b) \rm is straightforward.

\sxn{Monodromy Equal To One}

For simplicity, we first consider the special case where ${\cal M}$
is the identity (and $\delta{\cal M}{\cal M}^{-1}$  $=0$).
{}From eq. (1.5), this is a gauge invariant choice, as ${\cal M}
={\bf 1}$ defines the trivial adjoint orbit of $G$.  In
this case, $u$ and $v$ are singlevalued functions
on the circle, and the two form $\Omega$
seperates into left and right moving pieces, $\Omega^L$ and
$\Omega^R$.  From eqs. (2.4), we now have $\delta\Omega^L=
\delta\Omega^R= 0$.  It is therefore
sufficient to invert the two pieces seperately,
as the Poisson bracket between $u(x)$ and $v(y)$ will vanish.
Thus, in this case ${\cal G}_{\rho\sigma}(x,y) = 0$.
The symplectic structure involving $u(x)$ is determined
from the two form  $\Omega ^L$,
while the symplectic structure involving $v(x)$ is determined
from the two form  $\Omega ^R$.  We compute the former in a systematic
way in what follows.  The analysis is identical for the latter and
we shall not do it here.

For an $n$ dimensional group $G$ we can parametrize $u$ by $n$ angles
$\xi_\alpha
=\xi_\alpha(x)$, $\alpha=1,2,...,n$.
We need not specify the parametrization explicitly.  We next define
$\{\delta\xi_\alpha\}$
to be a basis of one forms.  In terms of this basis, Lie algebra
valued one forms
$ u^{-1}\delta u$
may be written
\begin{equation}
 u^{-1}\delta u = \delta\xi_\alpha  L_{\alpha\beta}T_\beta\quad.
\end{equation}
 $L_{\alpha\beta}$ defines a nonsingular
matrix which depends on the particular parametrization
 $u=u [\xi_\alpha ]$, as well as the coordinate $x$ on the circle.
\cite{marmo}  In addition, let $\{
{\delta\over \delta\xi_\alpha}
\}$ be a basis
of tangent vectors dual to
$\{\delta\xi_\alpha\}$; ie.
\begin{equation}
\delta\xi_\alpha(x)\Big|_
{\delta\over \delta\xi_\beta(y)}
=\delta_{\alpha\beta}\delta(x-y)\quad.
\end{equation}
$\{{\delta\over \delta\xi_\alpha}\}$
span the tangent vector space which we denote by $TQ_L$.
Then any vector $\chi$ in $TQ_L$ can be expanded according to
\begin{equation}
\chi=\int_0^\ell dx\;\chi_\alpha(x)
{\delta\over \delta\xi_\alpha(x)}\quad,
\end{equation}
$\chi_\alpha(x)$ denoting the components of $\chi$.
When the one form
$ u^{-1}\delta u $ is evaluated along the tangent vector
$\chi$ we get
\begin{equation}
[u^{-1}\delta u](x) \big|_\chi =
{\chi_\alpha(x)} L_{\alpha\beta}(x)T_\beta\quad.
\end{equation}
When the two form $\Omega^L$ is evaluated along directions
$\chi$ and $\psi$, we get
\begin{equation}
\Omega ^L\big|_{(\chi,\psi)} =
-{\kappa \over{ 4\pi}}
\int_0^\ell dx \;\chi_\alpha(x)L_{\alpha\beta}(x)\;
\partial_x\Bigl(\psi_\lambda(x)L_{\lambda\beta}(x)\Bigr)\quad,
\end{equation}
where $\psi_\alpha(x)$ are components of $\psi$
 and we have chosen the convention
  $Tr\; T_\alpha T_\beta =  \delta_{\alpha\beta}$.
Here the one form $ u^{-1}\delta u $ is singlevalued on the circle.
  From eq. (2.4),
for an arbitrary tangent vector $\chi$,
$\chi_\alpha(x)L_{\alpha\beta}(x)$ will also be singlevalued.
It then follows that $\Omega^L$ is antisymmetric, ie.
$\Omega ^L\big|_{(\chi,\psi)} =  -\Omega ^L\big|_{(\psi,\chi)}$.

 It is evident that the two form $\Omega ^L$ is degenerate; ie.
$\Omega ^L$ is zero when evaluated along certain \it zero vectors \rm
$\phi^{(\alpha)}$,
\begin{equation}
\Omega ^L\big|_{(\chi,\phi^{(\alpha)})} =  0\quad,
\end{equation}
for an arbitrary tangent vector $\chi$. ($\alpha$ is a degeneracy label.)
{}From eq. (3.5), the components $\phi^{(\alpha)}_\lambda(x)$
 of these zero vectors satisfy the requirement that
$\phi^{(\alpha)}_\lambda(x)L_{\lambda\beta}(x) = constant$.
There are $n$ independent solutions for $\phi^{(\alpha)}_\lambda(x)$
which correspond to $L^{-1}_{\alpha\lambda} (x)$.
More generally, the zero vectors are a linear combination of $L^{-1}$,
\begin{equation}
\phi^{(\alpha)}= C^{\alpha\beta}
\int_0^\ell dx\;
L^{-1}_{\beta\lambda}(x)
{\delta\over \delta\xi_\lambda(x)}\quad.
\end{equation}
We shall choose the constant matrix $C$ such that
$\phi^{(\alpha)}$ are orthonormal.  For this we define a scalar product
on the space of zero vectors
\begin{equation}
<\phi^{(\alpha)},
\phi^{(\beta)}> =
\int_0^\ell dx\;
\phi^{(\alpha)}_\lambda(x)
\phi^{(\beta)}_\lambda(x) = \delta_{\alpha\beta}\quad.
\end{equation}

Here, let us define $Q_L$ to be the space
spanned by $u(x)$.  It is a principal fibre bundle
with the structure group being $G$.  The action of $G$ on $Q_L$
is given in eq. (1.4).
The zero vectors which we have just obtained point in
the vertical directions of the principal bundle.
To see this, we can have them act on a point
$u(x)$ of $Q_L$.  From eq. (3.1),
$$  [ \phi^{(\alpha)}u  ] (x)=C^{\alpha\beta}u(x)T_\beta\; ,$$
and thus $i\epsilon_\alpha\phi^{(\alpha)}u$,
for infinitesimal $\epsilon_\alpha$,
corresponds to an infinitesimal change
in $u$ under a gauge transformation.  As gauge transformations are
associated with moving vertically along the fibre,
the $\phi^{(\alpha)}$'s must be vertical tangent vectors.

Let $TQ_L$ denote the tangent space of $Q_L$.
In order to invert $\Omega ^L$ and obtain the Poisson brackets
$\{,\}$ of the theory, we shall restrict it to acting on the
subspace of $TQ_L$ corresponding to the set of horizontal vectors.
  On this subspace
$\Omega ^L$ will be nonsingular.  This procedure, however,
 is not unique and requires making a gauge choice,
which we explain below.

Let us introduce a projection operator
 $P$, which in the basis  ${\delta\over \delta\xi_\alpha}$ has the form
\begin{equation}
P_{\alpha\beta}(x,y) =\delta_{\alpha\beta}\delta(x-y)
- \sum_\lambda \phi^{(\lambda)}_\alpha(x)
\phi^{(\lambda)}_\beta(y) \quad.
\end{equation}
$P$ projects out zero vectors, since
$\int_0^\ell dy\; P_{\alpha\beta}(x,y)\phi^{(\lambda)}_\beta(y) = 0$.
We now make the gauge choice that the remaining (horizontal) tangent
vectors are orthogonal to the zero vectors with respect to the
scalar product
 defined in eq. (3.8).  Then $P$ acts as the identity operator on
these tangent vectors.

 We can now use $P$ to define the Poisson brackets of any two angles
 $\xi_\alpha(x)$ and $\xi_\beta(y)$ as follows
\begin{equation}
\int_0^\ell dy\;\{\xi_\alpha(x),\xi_\beta(y)\}
\;\;\Omega ^L\Big|_{\big(
{\delta\over \delta\xi_\beta(y)},
{\delta\over \delta\xi_\gamma(z)}
\big)}=P_{\alpha\gamma}(x,z) \quad.
\end{equation}
By applying eq. (3.5), the right hand side of eq. (3.10) becomes
$$-{\kappa \over{ 4\pi}}
\int_0^\ell dy\;
\{\xi_\alpha(x),\xi_\beta(y)\}
\;L_{\beta\rho}(y)\;
\partial_y\Bigl(L_{\gamma\rho}(z)\delta(y-z)\Bigr)$$
\begin{equation}
= {\kappa \over{ 4\pi}} \partial_z\Bigl(
\{\xi_\alpha(x),\xi_\beta(z)\}
\;L_{\beta\rho}(z)\Bigr)\;L_{\gamma\rho}(z)\quad.
\end{equation}
Thus eq. (3.10) reduces to
\begin{equation}
 {\kappa \over{ 4\pi}}\partial_z\Bigl(
\{\xi_\alpha(x),\xi_\beta(z)\}
\;L_{\beta\rho}(z)\Bigr) = L_{\rho\gamma}^{-1}(z)
P_{\alpha\gamma}(x,z)\quad.
\end{equation}
Upon integration, we then find
$$ {\kappa \over{ 4\pi}}\{\xi_\alpha(x),\xi_\beta(z)\}\quad
=\quad\Bigl(L^{T\;-1}(x)L^{-1}(z)\Bigr)_{\alpha\beta}\;
\theta(z-x)\qquad $$
\begin{equation}
\qquad - \quad \Bigl(L^{T\;-1}(x)F(z)L^{-1}(z)\Bigr)_{\alpha\beta}
\quad + \quad \Bigl(f(x)L^{-1}(z)\Bigr)_{\alpha\beta}\quad,
\end{equation}
where $F(x)$ and $f(x)$ are matrix valued functions on the circle.
The former is given by the indefinite integral
\begin{equation}
F(z)=C^TC\int^z dy\;L^{-1}(y)L^{T\;-1}(y)\quad.
\end{equation}
To determine the latter, we require that the Poisson bracket
is antisymmetric,   \hfil \break
$\{\xi_\alpha(x),\xi_\beta(y)\}=$
$-\{\xi_\beta(y),\xi_\alpha(x)\}$.
{}From eq. (3.13), this leads to
\begin{equation}
-{\bf1}+F^T(x) + F(z)=f^T(z)L(z) +L^T(x)f(x)\quad,
\end{equation}
${\bf1}$ being the identity matrix.  The general solution to
eq. (3.15) is
\begin{equation}
f(x) = L^{T\;-1}(x)\Bigl(-{1\over 2}\;{\bf1} + F^T(x)
 + a_L\Bigr)\quad.
\end{equation}
$a_L$ is a constant (x-independent) antisymmetric matrix $a_L^T=-a_L$.
It can be viewed as a constant of integration.
After substituting this solution for $f(x)$ into eq. (3.13), we then find
\begin{equation}
\{\xi(x),\xi^T (y)\} = L^{T\;-1}(x){\cal F}(x,y)L^{-1}(y)\quad,
\end{equation}
\begin{equation}
-{\kappa \over{ 4\pi}}
{\cal F}(x,y)={\bf1}\;\epsilon (x-y) +F(y)-F^T(x)-a_L\quad,
\end{equation}
where $\epsilon(x)= {1\over 2}$ , for $x>0$,
and $\epsilon(x)= -{1\over 2}$ , for $x<0$.
${\cal F}$ satisfies the property of antisymmetry, ie.
${\cal F}^T(x,y)=-{\cal F}(y,x)$.
In addition, for ${\cal M} = 1$, it is singlevalued on the circle
with respect to both arguments $x$ and $y$.  To see this use
$F(\ell)=F(0)+{\bf 1}$, which follows from the othonormality
relation (3.8).

It is now easy to obtain the Poisson brackets between any two chiral
group elements $u(x)$ and $u(y)$.  For this note that from eq. (3.1), we
have the relation
$ L_{\alpha\rho}(x)T_\rho
\{\xi_\alpha(x),\;\}=u^{-1}(x)\{u(x),\;\}$.
Then if we multiply the Poisson bracket
$\{\xi_\alpha(x),\xi_\beta(y)\}$
in eq. (3.17) by the tensor product
$ L_{\alpha\rho}(x)L_{\beta\sigma}(y)
T_\rho \otimes T_\sigma$, we arrive at eq. (1.6),
with the function ${\cal F}$ given in eq. (3.18).
The Poisson bracket between any two group elements
$u$ evaluated at different points on the circle is thus seen to
depend on the way we parametrize the group elements in terms of
angles $\xi_\alpha$.  This is since $L$ and consequently
${\cal F}$ depends on the choice for $u=u [\xi_\alpha ]$.

Let us mention a few words about the arbitrary constant matrix $a_L$.
It is possible that $a_L$ gets determined by eq. (1.18),
which followed from the Jacobi identity.  Otherwise, its presence
would appear to indicate that we have not
fixed the gauge completely in the above solution for ${\cal F}$.
In any case, it
 does not however enter in the expressions for the
 Poisson brackets of known
observables of the theory, ie. the conserved currents.

As in ref. $ [11]$, we can recover the Kac-Moody algebra for the
currents of the theory starting with the Poisson brackets
between two chiral group elements.
 From eq. (1.6), we first easily obtain
the Poisson brackets involving $u^{-1}$
\begin{equation}
\{u^{-1}(x){}^\otimes_, u(y)\} =
-T_\rho u^{-1}(x)\otimes u(y)T_\sigma\;
{\cal F}_{\rho\sigma}(x,y)\quad,
\end{equation}
\begin{equation}
\{u^{-1}(x){}^\otimes_, u^{-1}(y)\} =
T_\rho u^{-1}(x)\otimes T_\sigma u^{-1}(y)\;
{\cal F}_{\rho\sigma}(x,y)\quad.
\end{equation}
Using these relations in addition to eq. (1.6), we can then
 compute the Poisson brackets between two left moving currents
 $J^L (x)=-\partial_x uu^{-1}(x)$,
\begin{equation}
\{J^L (x){}^\otimes_,J^L(y)\} =
u(x)T_\rho u^{-1}(x)\otimes u(y)T_\sigma u^{-1}(y)  \;
\partial_x \partial_y {\cal F}_{\rho\sigma}(x,y)\quad.
\end{equation}
These currents $J^L$ are gauge invariant.
Upon expanding them in terms of Lie algebra components
 $J^L (x)=$
${ {4\pi} \over{ i\kappa}} J^L_\alpha(x)T_\alpha$ and using
$\partial_x \partial_y {\cal F}_{\rho\sigma}(x,y) =$
$ {{4 \pi } \over \kappa}
\partial_x \delta(x-y)\delta_{\rho\sigma}$
, we then get the usual Kac-Moody algebra for $J^L_\alpha(x)$,
\begin{equation}
\{J^L_\alpha (x),J^L_\beta (y)\} =
c_{\alpha\beta\gamma} J^L_\gamma (x)\;\delta(x-y)\;-\;
{\kappa\over{4\pi}}\partial_x \delta(x-y)\delta_{\alpha\beta}\quad.
\end{equation}
In deriving eq. (3.22) we have assumed that
 $T_\alpha\otimes T_\alpha$ is adjoint invariant, ie.
$u(x)T_\alpha u^{-1}(x)\otimes u(x)T_\alpha u^{-1}(x) = $
$T_\alpha\otimes T_\alpha$.

\sxn{Arbitrary Monodromy  }

Next we consider the case of arbitrary monodromy ${\cal M}$.
For this let us first rename $\xi_\alpha$ by $\xi_{\alpha L}$.
As before, they coordinatize the left chiral group element $u$.
In addition, call $\xi_{\alpha R}$ the coordinates of the right
chiral group element $v$.  Now let $\{\delta\xi_{\alpha A}, A=L,R\}$
span the space of one forms associated with both $u$ and $v$.
In terms of them we can expand
$u^{-1}\delta u$ and $\delta v v^{-1}$ according to
\begin{equation}
 u^{-1}\delta u =
 \delta\xi_{\alpha L} L_{\alpha\beta}T_\beta\quad,\qquad
\delta vv^{-1} =  \delta\xi_{\alpha R} R_{\alpha\beta}T_\beta\quad.
\end{equation}
Like $L_{\alpha\beta}$, $R_{\alpha\beta}$ defines a nonsingular
matrix which depends on the particular parametrization of the
chiral group element, as well as the coordinate $x$ on the circle.
\cite{marmo}
We next define the tangent vector space $TQ$.  It is spanned by the set
of vectors $\{{\delta\over \delta\xi_{\alpha A}}\}$ which are dual to
$\{\delta\xi_{\alpha A}\}  $; ie.
\begin{equation}
\delta\xi_{\alpha A}(x)\Big|_ {\delta\over \delta\xi_{\beta B}(y)}
=\delta_{\alpha\beta}\delta_{AB}\delta(x-y)\quad.
\end{equation}
Then any vector $\chi$ in $TQ$ can be expanded according to
\begin{equation}
\chi=\int_0^\ell dx\;\chi_{\alpha A}(x)
{\delta\over \delta\xi_{\alpha A}(x)}\quad,
\end{equation}
$\chi_{\alpha A}(x)$ denoting the left and right components of $\chi$.
When the one form $u^{-1}\delta u$ is evaluated along the tangent vector
$\chi$ we obtain
\begin{equation}
[u^{-1}\delta u](x) \big|_\chi =   {\cal L}^{[\chi]}_\alpha (x)
T_\alpha\; ,\quad  {\cal L}^{[\chi]}_\alpha (x)
  =\chi_{\beta L}(x)L_{\beta \alpha}(x)\quad .
\end{equation}
When the one form
$ \delta v v^{-1} $ is evaluated along the tangent vector
$\chi$ we obtain
\begin{equation}
[\delta v v^{-1}](x) \big|_\chi =  {\cal R}^{[\chi]}_\alpha (x)
T_\alpha\; ,\quad  {\cal R}^{[\chi]}_\alpha (x)
 =\chi_{\beta R}(x)R_{\beta \alpha}(x)\quad .
\end{equation}

For arbitrary monodromy ${\cal M}$, we must consider all terms
in the symplectic two form eq. (2.1).  We can rewrite them as follows
$$\Omega = -{\kappa \over{ 8\pi}}
 \int_0^\ell dx\; Tr \biggl( [u^{-1}\delta u](x) \>
 \partial_x\Bigl( [u^{-1}\delta u](x) \Bigr)\;+ \;
[\delta  v v^{-1}](x)
 \partial_x\Bigl([\delta v v^{-1}](x) \Bigr)  $$
\begin{equation}
\qquad -\quad\partial_x \Bigl([u^{-1}\delta u](x) \; [
\delta v v^{-1}](\ell -x)\Bigr) \; \biggr) \quad .
\end{equation}
When the two form $\Omega$ is evaluated along directions
$\chi$ and $\psi$, we now get
$$\Omega \big|_{(\chi,\psi)} = -{\kappa \over{8\pi}}
\int_0^\ell dx \; \biggl( {\cal L}^{[\chi]}_\alpha (x)
\buildrel \leftrightarrow \over \partial _x
{\cal L}^{ [\psi]}_\alpha (x) \quad +\;
 {\cal R}^{ [\chi]}_\alpha (x)
\buildrel \leftrightarrow \over \partial _x
{\cal R}^{ [\psi]}_\alpha (x) $$
\begin{equation}
\qquad - \; \partial_x \Bigl( {\cal L}^{ [\chi]}_\alpha (x)
{\cal R}^{ [\psi]}_\alpha (\ell - x) \; - \;
 {\cal L}^{ [\psi]}_\alpha (x)
{\cal R}^{ [\chi]}_\alpha (\ell - x) \Bigl)\;\biggr)\quad,
\end{equation}
where $\buildrel \leftrightarrow \over \partial _x =
\buildrel \rightarrow \over \partial _x  -
\buildrel \leftarrow \over \partial _x  $.
Written in this form it is clear that $\Omega$ is antisymmetric, ie.
$\Omega \big|_{(\chi,\psi)} =  -\Omega \big|_{(\psi,\chi)}$.

For the previous case of
${\cal M}={\bf 1}$, $u$ and $v$ were singlevalued on
 the circle, and as a consequence so were ${\cal L}^{ [\chi]}$ and
${\cal R}^{ [\chi]}$.  This is not in general the case for arbitrary
$ {\cal M}$.  We shall instead fix the boundary conditions on
${\cal L}^{ [\chi]}$ and ${\cal R}^{ [\chi]}$ by demanding that
$\Omega \big|_{(\chi,\psi)} $ is differentiable in $\psi$.
(Similiarly, we fix the boundary conditions on
${\cal L}^{ [\psi]}$ and ${\cal R}^{ [\psi]}$ by requiring that
$\Omega \big|_{(\chi,\psi)} $ is differentiable in $\chi$.)
This leads to the conditions
\begin{equation}
{\cal L}^{ [\chi]}(\ell) + {\cal R}^{ [\chi]}(0)= 0\quad {\rm and}\quad
{\cal R}^{ [\chi]}(\ell) + {\cal L}^{ [\chi]}(0)=0\quad .
\end{equation}
With these restrictions we compute the variational derivatives of
$\Omega \big|_{(\chi,\psi)} $ with respect to the components of
$\psi$ to be
\begin{equation}
{{\partial\Omega \big|_{(\chi,\psi)}}\over{\partial\psi_{\beta L}(x)}}
= {\kappa\over{4\pi}}\;\partial_x
 {\cal L}^{ [\chi]}_\alpha (x)\;L_{\beta\alpha}(x)\;,\quad
{{\partial\Omega \big|_{(\chi,\psi)}}\over{\partial\psi_{\beta R}(x)}}
= {\kappa\over{4\pi}}\;\partial_x
 {\cal R}^{ [\chi]}_\alpha (x)\;R_{\beta\alpha}(x)\; .
\end{equation}
(The requirement of differentiability has also recently been shown
useful in analyzing the constraints and observables
of Chern-Simons theory.\cite{cs})

As with the case of ${\cal M}={\bf 1}$, the two form $\Omega$ is
singular and there exit zero vectors $\phi^{(\alpha)}$, where
\begin{equation}
\Omega\big|_{(\chi,\phi^{(\alpha)})} =  0\quad,
\end{equation}
for arbitrary tangent vectors $\chi$.
($\alpha$ is a degeneracy label.)  From eqs. (4.7) and (4.8),
it follows that ${\cal L}^{ [\phi^{(\alpha)}]} = -
{\cal R}^{ [\phi^{(\alpha)}]}= constant.$
n independent solutions for $\phi^{(\alpha)}$ are $(
\phi^{(\alpha)}_{\beta L}, \phi^{(\alpha)}_{\beta R})=$
$(L^{-1}_{\alpha \beta}, -
R^{-1}_{\alpha \beta}) $, or more generally, we can take
a linear combination of such solutions,
\begin{equation}
\phi^{(\alpha)}= C^{\alpha\beta}
\int_0^\ell dx\;   \biggl(L^{-1}_{\beta\lambda}(x)
{\delta\over \delta\xi_{\lambda L}(x)} -R^{-1}_{\beta\lambda}(x)
{\delta\over \delta\xi_{\lambda R}(x)}\biggr) \quad.
\end{equation}
We shall choose the constant matrix $C$ such that $\phi^{(\alpha)}$ are
again orthonormal, now with respect to the scalar product,
\begin{equation}
<\phi^{(\alpha)},\phi^{(\beta)}>=\int_0^\ell dx\;
\phi^{(\alpha)}_{\lambda A}(x)
\phi^{(\beta)}_{\lambda A}(x) = \delta_{\alpha\beta}\quad.
\end{equation}

As stated in the introduction, the space $Q$ spanned by $u(x)$ and $v(x)$
defines a principal fibre bundle with structure group $G$.
The action of $G$ on $Q$ is given by eq. (1.4), and once again,
the zero vectors point along
the vertical directions of the principal bundle.  The latter is true
since
$$  [ \phi^{(\alpha)}u ](x)=C^{\alpha\beta}u(x)T_\beta\quad {\rm and}
\quad  [\phi^{(\alpha)}v](x)=-C^{\alpha\beta}T_\beta v(x)\; $$
are associated with infinitesimal changes
in $u$ and $v$ under a gauge transformation.

Following the previous procedure, we define a projection operator
\begin{equation}
P_{(\alpha A)(\beta B)}(x,y)=\delta_{\alpha\beta}\delta_{AB}\delta(x-y)
- \sum_\lambda \phi^{(\lambda)}_{\alpha A}(x)
\phi^{(\lambda)}_{\beta B}(y)  \quad,
\end{equation}
for the purpose of projecting out the zero vectors.
It acts as the identity on horizontal vectors if we assume that the
latter are orthogonal to $\phi^{(\alpha)}$ with respect to scalar
product (4.12).  Again, this is a gauge choice.

We can then obtain the Poisson brackets between
any two angular variables $\xi_{\alpha A} (x)$ and $\xi_{\beta B}(y)$
by generalizing eq. (3.10) to
\begin{equation}
\int_0^\ell dy\;\{\xi_{\alpha A}(x),\xi_{\beta B}(y)\}
\;\;\Omega \Big|_{\big({\delta\over \delta\xi_{\beta B}(y)},\;\chi
\big)}= \int_0^\ell dy  \;
P_{(\alpha A)(\gamma D)}(x,y)\; \chi_{\gamma D} (y)   \quad,
\end{equation}
where $\chi$ is an arbitrary vector satisfying the boundary conditions
(4.8).

We analyze eq. (4.14) for the cases
 \it i) \rm $\chi={\delta\over \delta\xi_{\gamma L}(z)}$,
 and \it ii) \rm  $\chi = {\delta\over \delta\xi_{\gamma R}(z)}$,
where $z \not = 0\;  {\rm or }\;\ell$ in both of the cases.

 \it i) \rm $\chi={\delta\over \delta\xi_{\gamma L}(z)},
z \not = 0\;  {\rm or }\;\ell$.  In this case eq. (4.14) reduces to
\begin{equation}
 {\kappa \over{ 4\pi}}
\partial_z\Bigl(\{\xi_{\alpha A}(x),\xi_{\beta L}(z)\}
\;L_{\beta\rho}(z)\Bigr) = L_{\rho\gamma}^{-1}(z)  \;
P_{(\alpha A)(\gamma L)}(x,z)\quad,
\end{equation}
which is the generalization of eq. (3.12).  In fact, when we set $A=L$
it is identical to eq. (3.12)
(with $\xi_\alpha$ corresponding to $\xi_{\alpha L}$).
Therefore it leads to the previously found Poisson brackets
(3.17) and (3.18) between two left moving chiral modes.

If we instead set $A=R$, we obtain an equation involving
the Poisson brackets between one left moving and one
right moving chiral mode.  Then after integration, eq. (4.15) leads to
\begin{equation}
  {\kappa \over{ 4\pi}}\{\xi_{\alpha R}(x),\xi_{\beta L}(z)\}\;
=\; \Bigl(R^{T\;-1}(x)F(z)L^{-1}(z)\Bigr)_{\alpha\beta}
\; + \; \Bigl(\tilde f(x)L^{-1}(z)\Bigr)_{\alpha\beta}\quad,
\end{equation}
where $F(z)$ and $\tilde f(x)$ are matrix valued functions.
$F(z)$ was given in eq. (3.14), while we shall deduce the form (up to
a constant matrix) of $\tilde f(x)$ in what follows.

\it ii) \rm $\chi={\delta\over \delta\xi_{\gamma R}(z)},
z \not = 0\;  {\rm or }\;\ell$.
It is straightforward to repeat the above procedure for
this case.  Then instead of eq. (4.15), we have
\begin{equation}
 {\kappa \over{ 4\pi}}
\partial_z\Bigl(\{\xi_{\alpha A}(x),\xi_{\beta R}(z)\}
\;R_{\beta\rho}(z)\Bigr) = R_{\rho\gamma}^{-1}(z) \;
P_{(\alpha A)(\gamma R)}(x,z)\quad.
\end{equation}
After setting $A=R$, we can compute the Poisson brackets between
two right moving modes.  The results are obtained by simply
replacing "$L$" everywhere in eqs. (3.17) and (3.18) by "$R$", ie.
\begin{equation}
\{\xi_{\alpha R}(x),\xi_{\beta R} (y)\} =\Bigl( R^{T\;-1}(x)
{\cal H}(x,y)R^{-1}(y)\Bigr)_{\alpha \beta}\quad,
\end{equation}
where
\begin{equation}
-{\kappa \over{ 4\pi}}
{\cal H}(x,y)={\bf1}\;\epsilon (x-y) +H(y)-H^T(x)-a_R\quad.
\end{equation}
The function $H$ is the analogue of $F$ for the left movers,
\begin{equation}
H(z)=C^TC\int^z dy\;R^{-1}(y)R^{T\;-1}(y)\quad,
\end{equation}
and $a_R$ is a constant antisymmetric matrix.
Like ${\cal F}$, ${\cal H}$ satisfies the property of antisymmetry, ie.
${\cal H}^T(x,y)=-{\cal H}(y,x)$.

By taking index $A=L$ in eq. (4.17), we again
get a Poisson bracket between a left moving and right moving mode
\begin{equation}
  {\kappa \over{ 4\pi}}
\{\xi_{\alpha L}(x),\xi_{\beta R}(z)\}\;=\;
\Bigl(L^{T\;-1}(x)H(z)R^{-1}(z)\Bigr)_{\alpha\beta} \; + \;
\Bigl(\tilde h(x)R^{-1}(z)\Bigr)_{\alpha\beta}\quad,
\end{equation}
with $\tilde h(x)$ being a matrix valued function.
Upon comparing this Poisson bracket with (4.16), we see that
$\tilde h$ and $\tilde f$ must be satisfy the condition
\begin{equation}
-H^T(x) -F(z)=\tilde h^T(z)L(z) +R^T(x) \tilde f(x)\quad.
\end{equation}
The general solution for $\tilde h$ and $\tilde f$ is
\begin{equation}
\tilde h(x) = -L^{T\;-1}(x)\Bigl(F^T(x) - b\Bigr)\quad,
\end{equation}
\begin{equation}
\tilde f(x) = -R^{T\;-1}(x)\Bigl(H^T(x) + b^T\Bigr)\quad,
\end{equation}
$b$ being a constant matrix.  Then
\begin{equation}
\{\xi_{\alpha L}(x),\xi_{\beta R} (y)\}
=\Bigl( L^{T\;-1}(x) {\cal G}(x,y)R^{-1}(y)\Bigr)_{\alpha \beta}\quad,
\end{equation}
where
\begin{equation}
-{\kappa \over{ 4\pi}} {\cal G}(x,y)= -H(y)+F^T(x)-b\quad.
\end{equation}

The Poisson brackets between all angular variables $\xi_{\alpha A}(x)$
are now specified up to the constant matrices $a_L, a_R$ and $b$.
To get the Poisson brackets between chiral group elements $u(x)$
and $v(x)$, multiply
$\{\xi_{\alpha L}(x),\xi_{\beta R}(y)\}$ by the tensor product
$ L_{\alpha\rho}(x)R_{\beta\sigma}(y) T_\rho\otimes T_\sigma$, and
$\{\xi_{\alpha R}(x),\xi_{\beta R}(y)\}$ by the tensor product
$ R_{\alpha\rho}(x) R_{\beta\sigma}(y) T_\rho\otimes
T_\sigma$.  The results are eqs. (1.7) and (1.8), with ${\cal H}$
and ${\cal G}$ defined in eqs. (4.19) and (4.26).  In addition, eq. (1.6)
once again follows from (3.17), with ${\cal F}$ defined in (3.18).

\sxn{Concluding Remarks}

Here we give some concluding remarks concerning the solutions we
have found in sections 3 and 4.

The constant matrices $a_L$, $a_R$ and $b$
 appearing in the solutions for ${\cal F}$, ${\cal G}$ and ${\cal H}$
are not independent.
This is due to boundary conditions on $u$ and $v$.  To show
this, let us first fix the lower limits in the defining
integrals for $F(x)$ and $H(x)$ in eqs.
(3.14) and (4.20) to be $0$.  Then
from the orthonormality property of zero vectors (4.12), we have
\begin{equation}
F(\ell) + H(\ell)= {\bf 1}  \quad.
\end{equation}
We next compute the Poisson brackets between ${\cal M}$ and
$v(x)$.  This can be done either by setting ${\cal M} = u^{-1}(0)
u(\ell)$ or ${\cal M} = v(\ell)v^{-1}(0)$. For the former we get
\begin{equation}
\{{\cal M}{}^\otimes_, v(y)\} =
\Bigl({\cal M}T_\rho\; {\cal G}_{\rho\sigma}(\ell,y)\;-\;
T_\rho  {\cal M}\;{\cal G}_{\rho\sigma}(0,y)\Bigr)
\otimes T_\sigma v(y)\quad ,
\end{equation}
while for the latter
\begin{equation}
\{{\cal M}{}^\otimes_, v(y)\} =
\Bigl(-{\cal M}T_\rho\; {\cal H}_{\rho\sigma}(0,y)\;+\;
T_\rho  {\cal M}\;{\cal H}_{\rho\sigma}(\ell,y)\Bigr)
\otimes T_\sigma v(y)\quad .
\end{equation}
These two formulae are consistent only if
\begin{equation}
b=-a_R - H^T(\ell)+ {1\over 2}{\bf1}\quad.
\end{equation}

We can get a relation between $b$ and $a_L$ by repeating the above
procedure for computing the Poisson brackets
between ${\cal M}$ and $u(x)$.  This gives
$$\{u(x){}^\otimes_,{\cal M}\} = u(x)T_\sigma\otimes
\Bigl({\cal M}T_\rho\; {\cal F}_{\sigma\rho}(x,\ell)\;-\;
T_\rho  {\cal M}\;{\cal F}_{\sigma\rho}(x,0)\Bigr)$$
\begin{equation}
\qquad\qquad =u(x)T_\sigma\otimes
\Bigl(-{\cal M}T_\rho \;{\cal G}_{\sigma\rho}(x,0)\;+\;
T_\rho  {\cal M}\;{\cal G}_{\sigma\rho}(x,\ell)\Bigr)\quad.
\end{equation}
Consistency now requires
\begin{equation}
b=-a_L - H(\ell)+ {1\over 2}{\bf1}\quad.
\end{equation}
So $a_L$ and $a_R$ are related by
$a_L -a_R =-H(\ell)+H^T(\ell)$, and there is only one independent
constant matrix in our solution.  Like in section 3, it is possible
that the remaining constant can get fixed by imposing Jacobi
identities.  If this is not the case, it would appear to
indicate that we have not fixed the gauge completely in the above
computations.

As we mentioned in the introduction, the meaning of the
quantum commutator analogues of the
Poisson bracket relations (1.6-8) is not transparant, because $u$
and $v$ are not the gauge invariant observables of the theory.  In
addition, of course, there are operator ordering ambiguities
encountered in quantization.  It would be of interest to understand
under what conditions the resulting commutation relations form a
quantum exchange algebra.

With regards to the gauge invariant observables of the theory, along
with the left moving current $J^L (x)=-\partial_x uu^{-1}(x)$
discussed in section 3, there is the right moving current
 $J^R (x)= v^{-1}\partial_x v(x) $.  $J^L$ and $J^R$ satisfy the Poisson
bracket relations
\begin{equation}
\{J^L (x){}^\otimes_,J^R(y)\} =  -
u(x)T_\rho u^{-1}(x)\otimes v^{-1}(y)T_\sigma v(y)  \;
\partial_x \partial_y {\cal G}_{\rho\sigma}(x,y)\quad,
\end{equation}
\begin{equation}
\{J^R (x){}^\otimes_,J^R(y)\} =
v^{-1}(x)T_\rho v(x)\otimes v^{-1}(y)T_\sigma v(y)  \;
\partial_x \partial_y {\cal H}_{\rho\sigma}(x,y)\quad,
\end{equation}
as well as (3.21).
But $\partial_x \partial_y {\cal G}_{\rho\sigma}(x,y) =0$, from
eq. (4.26), so the right hand side of eq. (5.7) vanishes.
Using $\partial_x \partial_y {\cal H}_{\rho\sigma}(x,y) =$
$ {{4 \pi } \over \kappa}
\partial_x \delta(x-y)\delta_{\rho\sigma}$,
and expanding $J^R$ in terms of Lie algebra components , we see
that eq. (5.8) corresponds to
the Kac-Moody algebra for the right moving current.
It is identical in form to that for $J^L$, ie. eq. (3.22).

Finally, we can compute Poisson brackets between gauge invariant
quantities formed from the monodromy matrix ${\cal M}$.  These are
just the adjoint invariants, such as
$det{\cal M}$ and $Tr{\cal M}^n$.  Note that ${\cal M}$ can not be
obtained from the conserved currents since it is not gauge invariant.
The Poisson brackets of ${\cal M}$ with itself can be gotten
directly from eq. (5.2).  We find
$$\{ {\cal M} {}^\otimes_, {\cal M} \}\; =  \;
{\cal M}T_\rho \otimes T_\sigma {\cal M} \;
{\cal G}_{\rho\sigma}(\ell,\ell)  \;+\;
T_\rho{\cal M} \otimes{\cal M} T_\sigma\;{\cal G}_{\rho\sigma}(0,0)
\qquad$$
\begin{equation}
\qquad - \;
T_\rho {\cal M} \otimes T_\sigma{\cal M} \;{\cal G}_{\rho\sigma}(0,\ell)
\;-\;
{\cal M}T_\rho \otimes{\cal M} T_\sigma\;{\cal G}_{\rho\sigma}(\ell,0)\;.
\end{equation}
Using this relation and the solution for ${\cal G}$ given in eq. (4.26),
we then see that the Poisson brackets between quantities such as
$det{\cal M}$ and $Tr{\cal M}^n$ vanish.  The adjoint invariants have
zero Poisson brackets with themselves, and consequently, the
associated quantum operators are simultaneously diagonalizable.

{\bf Acknowledgements}

We have been supported during the course of this work as follows: 1)
G. B. by the Department of Energy, USA, under contract number D
E-FG-02-85ER-40231, and A. Stern
by the Department of Energy, USA under
contract number DE-FG-05-84ER-40141;
2) A. Stern by INFN, Italy [at Dipartimento di
Scienze Fisiche, Universit{\`a} di Napoli]; 3) G. B. and A. Simoni
 by the Dipartimento
di Scienze Fisiche, Universit{\`a} di Napoli; 4) P. S. by the Swedish
National Science Research Council under contract number F-FU 8230-303.
A. Stern wishes to thank the
group in Naples, Giuseppe Marmo, in particular, and the Institute for
Theoretical Physics in G\"oteborg, B.-S. Skagerstam, in particular,
for their hospitality while this work was in progress.
We also wish to thank A. P. Balachandran, I. Bengtsson, G. Marmo,
B.-S. Skagerstam and F. Zaccaria for very helpful comments.

\end{document}